\documentstyle[12pt,worldsci]{article}
\begin{document}
\def\z2{\ifmmode Z_2\else $Z_2$\fi}
\def\ie{{\it i.e.},}
\def\eg{{\it e.g.},}
\def\etc{{\it etc}}
\def\etal{{\it et al.}}
\def\ibid{{\it ibid}.}
\def\to{\rightarrow}
\def\epem{\ifmmode e^+e^-\else $e^+e^-$\fi}
\def\Re{{\cal R \mskip-4mu \lower.1ex \hbox{\it e}\,}}
\def\Im{{\cal I \mskip-5mu \lower.1ex \hbox{\it m}\,}}
\pagestyle{empty}
\setlength{\baselineskip}{2.6ex}

\title{{\bf EXTENDING THE KINEMATIC RANGE FOR $W_R$ SEARCHES IN $e^-e^-$
COLLISIONS AT THE NLC}}
\author{Thomas G.~Rizzo\\
%\vspace{0.3cm}
\vspace{0.3cm}
{\em Stanford Linear Accelerator Center, Stanford University, Stanford, CA
94309, USA}}
\maketitle

\begin{center}
\parbox{13.0cm}
{\begin{center} ABSTRACT \end{center}
{\small\hspace*{0.3cm}

While the much discussed lepton-number violating process $e^-e^-\rightarrow
W_R^-W_R^-$ provides an excellent probe of both the Majorana nature of the
right-handed neutrino and the symmetry breaking sector of the Left-Right
Symmetric Model, it is likely that $W_R$'s are too massive to be pair produced
at the NLC with $\sqrt {s}$ in the 1-1.5 TeV range. We are thus lead to
consider the single $W_R$ production process $e^-e^-\rightarrow W_R^-(W_R^-)^*
\rightarrow W_R^-+jj$ in order to expand the collider's kinematic reach. After
pointing out that $W_R$'s with masses of order 1 TeV may be missed by future
hadron collider searches, we demonstrate that this three-body process
possesses a significant cross section, of order several fb, at the NLC with
$\sqrt {s}$ in the range above. The angular distribution of the produced
$W_R$'s is shown to be essentially flat and the potential backgrounds from
standard model processes are shown to be small.}}

\end{center}

The possibility of producing like sign pairs of $W$ bosons in $e^-e^-$
collisions has been discussed for some time{\cite {1}}. Such a process, if it
exists, signals the existence of new $|\Delta L|=2$ interactions which may
manifest themselves as Majorana masses for neutrinos. Within the Standard
Model(SM) gauge group, it is difficult to generate a large cross section for
this reaction while simultaneously satisfying the constraint of tree-level
unitarity at large values of the center of mass energy, $s$, and the bounds on
the effective neutrino mass arising from the lack of observation of
neutrinoless double beta decay. These difficulties can be easily circumvented
by extending the gauge group to that of the Left-Right Symmetric
Model(LRM){\cite {2}}
and considering instead the reaction $e^-e^- \to W_R^-W_R^-$, where $W_R$ is
the right-handed charged gauge boson. This process occurs quite naturally in
the LRM as a result of the see-saw mechanism used to generate small masses for
the ordinary `left-handed' neutrinos.

The amplitude for $e^-e^- \to W_R^-W_R^-$ gets both
$t-$ and $u-$channel contributions from the exchange of heavy `right-handed'
neutrinos($N$), with mass $M_N$, as well as an $s-$channel contribution from
the exchange of a doubly-charged Higgs boson($\Delta$), with mass $M_\Delta$.
(Any mixing between the SM $W$ and $W_R$ will be neglected in what follows.)
Since the $e^-e^-\Delta$ coupling is proportional to $M_N$ and the $e^-NW_R$
coupling is chiral, the total amplitude is found to be proportional to $M_N$.
Thus, as the Majorana mass of $N$ vanishes so does the amplitude, which is
just what we would expect since it is this Majorana mass term which generates
the $|\Delta L|=2$ interaction.  At NLC energies, \ie  $\sqrt {s}=0.5-1.5$ TeV,
the cross section for  $e^-e^- \to W_R^-W_R^-$ is quite large, of order a few
$pb$, fairly sensitive to the values of $M_N$ and $M_\Delta$, and has
a rather flat angular distribution. The $s-$channel $\Delta$ may appear as a
resonance depending upon the value of $\sqrt {s}$. Unfortunately, the `reach'
is rather limited since we are restricted to $W_R$ masses less than
$\sqrt {s}/2$ and there are substantial reasons to believe{\cite {3}} that
$W_R$'s are relatively heavy with masses $M_R\geq 0.5$ TeV. It is reasonable
to contemplate
that $W_R$ pair production may not be kinematically accessible at these
center of mass energies. This forces us to consider{\cite {4}} the
possibility of {\it singly} producing $W_R$'s via the reaction
$e^-e^- \to W_R^-(W_R^-)^*\to W_R^-jj$. We limit ourselves to this $jj$
mode to allow for the possibility that $M_N>M_R$ in which case $W_R$ can only
decay to $jj$ barring the existence of exotics. It is interesting to
note that all collider searches for $W_R$ rely on it's leptonic decay as a
trigger; if $M_N>M_R$ it is quite possible that $W_R$'s may not be observable
at the Tevatron or LHC{\cite {5}} and may be missed until the NLC turns on.

Of course, allowing one of the $W_R$'s to be off-shell we are forced to pay
the price of an additional gauge coupling as well as three-body phase space.
This results in a substantial reduction in the cross section from the on-shell
case to the level of a few $fb$. This implies machine luminosities in the range
of ${\cal L}=100-200 fb^{-1}$ are required to make use of this channel.
The complete expression for the cross section is given in Ref. 4.
The total event rates for $e^-e^- \to W_R^-(W_R^-)^*\to W_R^-jj$ are
found in Figs.~1 and 2, in which we have set $\kappa=1$ and scaled by an
integrated luminosity of $100 fb^{-1}$. ($\kappa=g_R/g_L$ is the ratio of the
two gauge couplings in the LRM.) Fig.~1a shows the number of expected
$W_R+jj$ events, as a function of $M_R$, at a $\sqrt {s}=1$ TeV $e^-e^-$
collider for different choices of $M_N$ and $M_{\Delta}$. The results are seen
to be quite sensitive to the values of these mass parameters even when $M_R$
is fixed. In Fig.~1b(c), we fix $M_R=700$ GeV and plot the event rate
as a function of $M_N(M_{\Delta})$ for various values of $M_{\Delta}(M_N)$.
Typically, we see event rates of order several hundred/yr except near the
$\Delta$ resonance (where very large rates are obtained) or when $M_N$ is
small (as the cross section vanishes for massless $N$ since it probes the $N$'s
Majorana nature). Increasing $\sqrt {s}$ to 1.5 TeV, as shown in Fig.~2a, we
see substantial cross sections are obtainable even assuming $W_R$'s in
the 1-1.2 TeV mass
range for some parameter choices. Fixing $M_R=1$ TeV in Figs.~2b and c, we
again see reasonable event rates for most choices of $M_N$ and $M_{\Delta}$
assuming $\sqrt {s}=1.5$ TeV. The exact rate is, however, a sensitive probe of
both the $N$ and $\Delta$ masses.
For most choices of the input masses we obtain extremely
flat distributions, however, when $N$ is light a significant angular
dependence is observed. This is simply a result of the $t-$ and $u-$ channel
poles which develop as $M_N$ tends to zero. Of course, small $M_N$ also leads
to a small cross section, as shown in Figs.~1 and 2, as might be expected
since the matrix element vanishes in this massless limit.

Potential backgrounds to the process $e^-e^- \to W_R^-(W_R^-)^* \to W_R^- +jj$
at the NLC are easily controlled and/or removed. For example, there may be
some contamination from the SM process $e^-e^- \to W_L^-W_L^- \nu \nu$, but
this can be easily eliminated by using missing energy cuts and demanding that
the $W_R$ final state be reconstructed from either the $jj$ or $eN \to eejj$
decay modes. (Since the on-shell $W_R$ decays to either $jj$ or $eN \to eejj$
there is no missing energy in the signal process.)
In addition, with polarized beams, we can take advantage of the
fact that $W_R$ couples via right-handed currents while any SM background must
arise only via left-handed currents. Within the LRM itself a possible
background could arise from a similar lepton-number conserving processes such
as  $e^-e^- \to W_R^-W_R^- NN$.
Even if such a final state could be produced, in comparison to the process we
are considering, the subsequent $N$ decays would lead to a final state with too
many charged leptons and/or jets.

In this talk the following points have been addressed:
($i$) While $e^-e^- \to W_R^-W_R^-$ is an excellent probe of both the Majorana
nature of $N$ and the symmetry breaking sector of the Left-Right Symmetric
Model, it is more than likely that $W_R$'s are too massive to be pair
produced at the NLC if $\sqrt {s}=1-1.5$ TeV forcing us to consider the
production of a single on-shell $W_R$ via the process
$e^-e^- \to W_R^-(W_R^-)^* \to W_R^- +jj$.
($ii$) Since the pair of on-shell $W_R$'s cross section was generally very
large, we would expect that the single $W_R$ rate would be significant if
integrated luminosities in the $100 fb^{-1}$ range were available. From the
explicit calculations we found that these expectations were realized for
most of the model parameter space with cross sections of order
$1-10 fb^{-1}$.
($iii$) For values of the input parameters that lead to significant rates,
the $W_R$ angular distribution was found to be rather flat implying that
angular cuts will not significantly reduce the cross sections. The rates
themselves were found to be quite sensitive to the particular values of the
masses of $N$ and $\Delta$. Masses for both these particles beyond the
kinematic reach of the NLC were found to be probed by the single $W_R$
production process.

$e^-e^-$ collisions allow us to probe the Majorana nature of the heavy
neutrinos in the LRM even when they are too massive to be directly produced.

\vspace{1.0cm}
%
%%%%%%%%%%%%%%%%%%%%%%%%%%%%%%%%%%%%%%%%%%%%%%%%%%%%%%%
\def\MPL #1 #2 #3 {Mod.~Phys.~Lett.~{\bf#1},\ #2 (#3)}
\def\NPB #1 #2 #3 {Nucl.~Phys.~{\bf#1},\ #2 (#3)}
\def\PLB #1 #2 #3 {Phys.~Lett.~{\bf#1},\ #2 (#3)}
\def\PR #1 #2 #3 {Phys.~Rep.~{\bf#1},\ #2 (#3)}
\def\PRD #1 #2 #3 {Phys.~Rev.~{\bf#1},\ #2 (#3)}
\def\PRL #1 #2 #3 {Phys.~Rev.~Lett.~{\bf#1},\ #2 (#3)}
\def\RMP #1 #2 #3 {Rev.~Mod.~Phys.~{\bf#1},\ #2 (#3)}
\def\ZP #1 #2 #3 {Z.~Phys.~{\bf#1},\ #2 (#3)}
\def\IJMP #1 #2 #3 {Int.~J.~Mod.~Phys.~{\bf#1},\ #2 (#3)}
\bibliographystyle{unsrt}

\begin{thebibliography}{99}
\bibitem{1}
T.G.~Rizzo, {\em Phys.~Lett.} {\bf B116}, 23 (1982);
D.~London, G.~Belanger, and J.N.~Ng, {\em Phys.~Lett.} {\bf B188},
155 (1987); J.~Maalampi, A.~Pietil\" a, and J.~Vuori, {\em Phys.~Lett.}
{\bf B297}, 327 (1992) and Turku University report FL-R9 (1992); M.P.~Worah,
Enrico Fermi Institute report EFI 92-65 (1992);
C.A.~Heusch and P.~Minkowski, CERN report CERN-TH-6606-92 (1993); see also
T.G.\ Rizzo in, {\it Proceedings of the Workshop on Physics and Experiments
with Linear $e^+e^-$ Colliders}, Waikoloa, Hawaii, April 1993,
edited by F.A.\ Harris \etal, (World Scientific, Singapore, 1993).
\bibitem{2}
For a review of the LRM and original references, see R.N. Mohapatra, {\it
Unification and Supersymmetry}, (Springer, New York, 1986).
\bibitem{3}
P.\ Langacker and S.U.\ Sankar, \PRD D40 1569 1989  ;
F.\ Abe \etal, CDF Collaboration, \PRL 67 2609 1991 ~and \PRL 68 1464 1992 ;
{}~see also the CDF and D0 Collaboration talks given at the {\it 9th Topical
Workshop on Proton-Antiproton Collider Physics}, Tsukuba, Japan, October 1993.
For an overview, see T.G.\ Rizzo, \PRD D50 325 1994 .
\bibitem{4}
T.G.\ Rizzo, SLAC report SLAC-PUB-6475, 1994.
\bibitem{5}
A.\ Datta, M.\ Guchait, and D.P.\ Roy, \PRD D47 961 1993  ;
D.\ Gingrich \etal, ATLAS Collaboration Letter of Intent, CERN report LHCC/I2,
(1992); A.\ Henriques and L.\ Poggioli, ATLAS Collaboration Note PHYS-NO-010,
(1992); T.G.\ Rizzo, \PRD D48 4236 1993  .

%
\end{thebibliography}

%$   figure  captions
{%\small
%\vspace*{2.00in}
\noindent
Fig.~1:
Event rates per $100 fb^{-1}$ for $W_R+jj$ production at a
1 TeV $e^-e^-$ collider assuming $\kappa=1$ (a) as a function of
$M_R$ for $M_N=M_{\Delta}=1$ TeV
(dots), $M_{\Delta}=1.2$ TeV and $M_N=0.4$ TeV (dashes), $M_{\Delta}=0.3$
and $M_N=0.1$ TeV (dash-dots), $M_{\Delta}=2,~M_N=0.6$ TeV (solid), or
$M_{\Delta}=1.8$ and $M_N=0.6$ TeV (square dots); (b) with $M_R=700$ GeV fixed
as a function of $M_N$ for $M_{\Delta}=0.3(0.6, 1.2, 1.5, 2)$ TeV
corresponding to the dotted(dashed, dash-dotted, solid, square-dotted) curve;
(c) as a function of $M_{\Delta}$ for $M_N=0.2(0.5, 0.8, 1.2, 1.5)$ TeV
corresponding to the dotted(dashed, dash-dotted, solid, square-dotted) curve.

\medskip

%\vspace*{2.15in}
\noindent
Fig.~2: Same as Fig.~1, but for a
1.5 TeV $e^-e^-$ collider. In (b) and (c), a $W_R$ mass of 1 TeV is assumed.

\medskip

 }

\end{document}